\documentclass[%
reprint,
superscriptaddress,
 amsmath,amssymb,
 aps,
pra,
]{revtex4-2}

\usepackage{siunitx}
\usepackage{float}
\usepackage{graphicx}
\usepackage{dcolumn}
\usepackage{bm}
\usepackage{xcolor}
\usepackage{color,soul}

\usepackage[dvipsnames]{xcolor}

\DeclareRobustCommand{\scottl}[1]{ \colorbox{LimeGreen!30}{(scott:) \href{#1}{link}} }

\begin{document}

\preprint{APS/123-QED}

\title{A scalable infrastructure for strontium optical clocks with integrated photonics}

\author{Zheng Luo}
\affiliation{Time and Frequency Division, National Institute of Standards and Technology, Boulder, Colorado, USA}
\affiliation{Department of Physics, University of Colorado, Boulder, Colorado, USA}

\author{Travis C. Briles}
\affiliation{Time and Frequency Division, National Institute of Standards and Technology, Boulder, Colorado, USA}

\author{Zachary L. Newman}
\affiliation{Time and Frequency Division, National Institute of Standards and Technology, Boulder, Colorado, USA}
\affiliation{Octave Photonics, Louisville, Colorado, USA}

\author{Aidan R. Jones}
\affiliation{Time and Frequency Division, National Institute of Standards and Technology, Boulder, Colorado, USA}
\affiliation{Department of Physics, University of Colorado, Boulder, Colorado, USA}

\author{Andrew R. Ferdinand} 
\author{Sindhu Jammi}
\affiliation{Time and Frequency Division, National Institute of Standards and Technology, Boulder, Colorado, USA}
\affiliation{Department of Physics, University of Colorado, Boulder, Colorado, USA}

\author{Grisha Spektor}
\affiliation{Time and Frequency Division, National Institute of Standards and Technology, Boulder, Colorado, USA}
\affiliation{Department of Physics, University of Colorado, Boulder, Colorado, USA}

\author{David R. Carlson}
\affiliation{Octave Photonics, Louisville, Colorado, USA}

\author{Akash Rakholia}
\author{Dan Sheredy}
\author{Parth Patel}
\author{Martin M. Boyd} 
\affiliation{Vector Atomic, Inc. Pleasanton, California, USA}

\author{Chad Ropp} 
\author{Daron Westly}
\author{Vladimir A. Aksyuk}
\affiliation{Microsystems and Nanotechnology Division, National Institute of Standards and Technology, Gaithersburg, Maryland, USA}

\author{Wenqi Zhu}
\author{Junyeob Song}
\author{Amit Agrawal} 
\affiliation{Microsystems and Nanotechnology Division, National Institute of Standards and Technology, Gaithersburg, Maryland, USA}

\author{Scott B. Papp}
 \email{scott.papp@nist.gov}
\affiliation{Time and Frequency Division, National Institute of Standards and Technology, Boulder, Colorado, USA}
\affiliation{Department of Physics, University of Colorado, Boulder, Colorado, USA}
\date{\today}

\begin{abstract}
Optical atomic clocks provide exceptionally accurate and precise signals for timekeeping and precision measurements, but they require high-power, free-space laser configurations that limit scalability. We introduce and explore a scalable infrastructure for strontium (Sr) optical-lattice clocks that incorporates co-design of atomic-beam slowing and a magneto-optical trap (MOT) from an effusion source, generation of complex, three-dimensional free-space laser configurations with a photonic integrated circuit (PIC) and metasurface (MS) optics, and laser stabilization to a frequency-comb supercontinuum generated with integrated nonlinear photonics. With these elements, we realize MOTs of all stable strontium isotopes ($^{84}$Sr, $^{86}$Sr, $^{87}$Sr, $^{88}$Sr) with populations commensurate with natural abundances, demonstrating precise beam control and robustness. Access to laser-cooled alkaline-earth atoms with scalable integrated photonics enables system engineering for optical clocks, quantum sensing, and quantum information, and our  experiments demonstrate extensible technologies that advance toward a Sr optical clock largely free of bulk optics.
\end{abstract}

\maketitle

\section{Introduction}
Coherence is an essential resource in modern science and technology that underpins high-speed optical communication \cite{kikuchi_fundamentals_2016}, microwave wireless communications \cite{xiao_millimeter_2017}, and physical sensing and network synchronization \cite{clivati_common-clock_2020, lu_distributed_2019}. Enabled by decades of development in digital electronics, microwave signals derive from low-phase-noise oscillators that are frequency stabilized to atomic clocks. A more recent development is the optical-lattice clock, which exploits the extremely narrow transitions of alkaline-earth atoms confined in a wavelength-scale optical potential \cite{Ludlow2015OpticalClocks}. Moreover, optical clocks offer a potential route for redefinition of the SI second through their accuracy and precision \cite{riehle_towards_2015, dimarcq_roadmap_2024}. This raises the goal of system integration in optical clocks and the use of photonic integrated circuit (PIC) technologies \cite{Spektor2023UniversalPhotonics, brodnik_monolithic_2025}. Useful PIC designs that operatively link light-atom interactions could make optical coherence available in applications, such as quantum information \cite{reichardt_fault-tolerant_2025}, quantum sensing \cite{takamoto2020test}, high-capacity data links \cite{Brodnik2021OpticallyLasers, pirmoradi_integrated_2025}, and novel optical computing architectures \cite{jin_kerr_2025}.

Optical clocks are composed of subsystems from narrow-linewidth, high-power lasers to pristine vacuum and electromagnetic-field environments \cite{Riehle2017}. Due to their complexity, state-of-the-art optical clocks are maintained in laboratories. Transportable optical clocks have been developed with reduced size and weight and enhanced robustness, promoting metrological innovation and exploration of applications like monitoring changes in the gravitational potential caused by volcano activity \cite{10.1093/gji/ggv246}. Strontium lattice clocks have been packaged to $\sim1$ m$^3$ volume and demonstrated with systematic fractional frequency uncertainty of $2\times10^{-17}$ \cite{PhysRevA.98.053443}, and a transportable calcium ion clock was used for a geopotential measurement \cite{PhysRevA.102.050802}. Two strontium optical lattice clocks with systematic uncertainty at the $10^{-18}$ level have been transported to a 450 m tower to compare their signals and test general relativity \cite{takamoto2020test}. Optical clocks based on lower complexity atomic structure have also been explored, including Rb two-photon \cite{PRA2018compactRb, Lemke2022MeasurementDays} and iodine \cite{Doringshoff2019IodineRocket,Sanjuan2021SimultaneousReference, Roslund2024OpticalSea}.

\begin{figure*}[tbp] \centering
\includegraphics[width=0.95\linewidth]{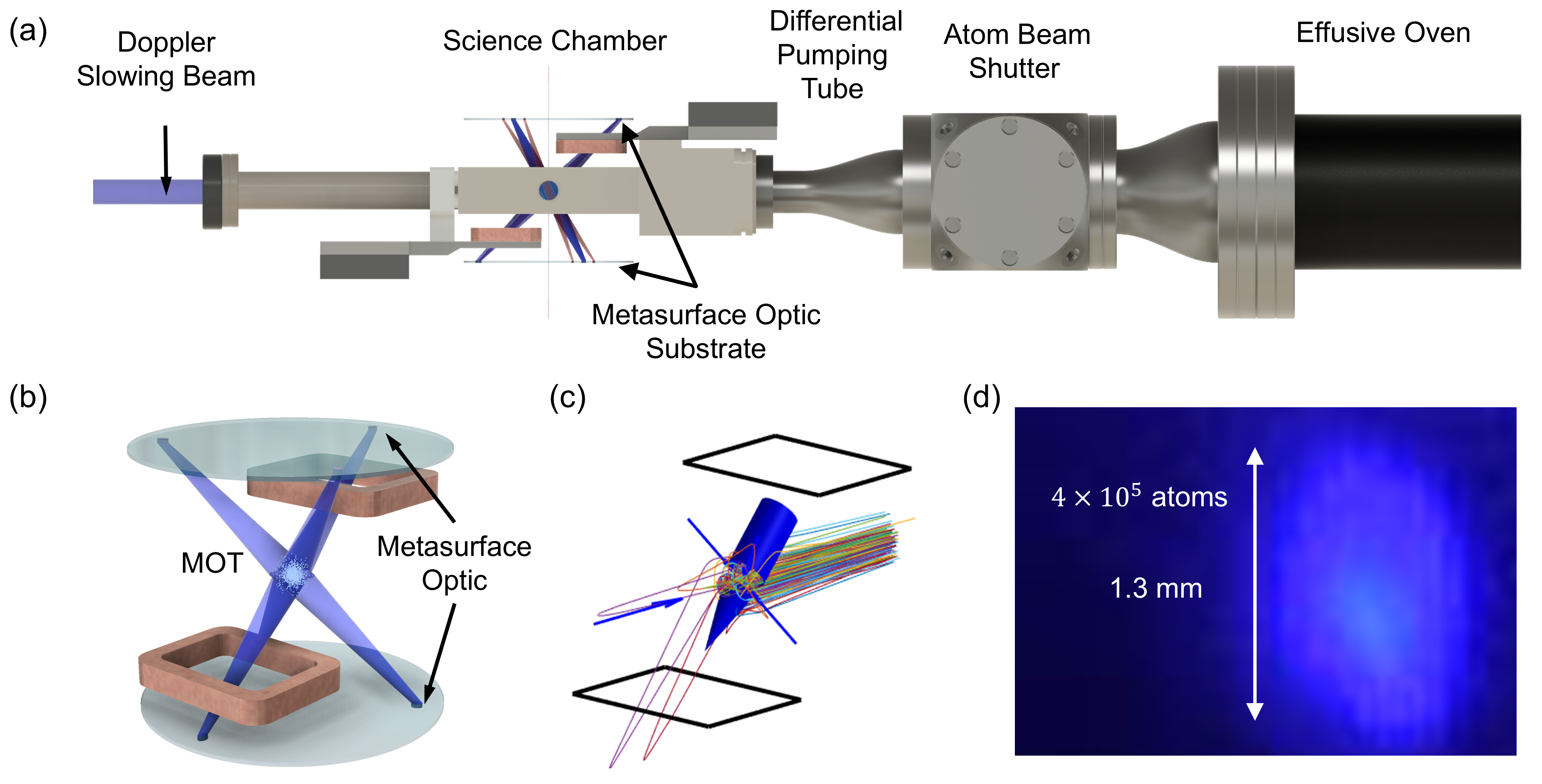}
\caption{\label{fig:fig1}\textbf{Scalable infrastructure for optical lattice clocks.} \textbf{(a)} Our Sr system includes a commercial effusive oven, mechanical feed through atomic beam shutter, differential pumping tube, atom-photonic interface such as metasurface chips, science chamber, and Doppler slowing beam introduced through a heated sapphire window. \textbf{(b)} Illustration of a customizable arrangement of multifunctional metasurface optics for MOT beam generation \textbf{(c)} Monte Carlo simulations of laser cooled atom trajectories of the planar MOT beam geometry. \textbf{(d)} Image of a realized  $4\times10^5$ atom $^{87}$Sr MOT.}
\end{figure*}

Given the potential impact of optical clocks, there has been substantial effort at enhancing their system integration, particularly by development of photonics devices. Some optical-clock technologies like self-referenced, phase-locked frequency combs have crosscutting applications, hence combs have reached commercialization and chip-integration \cite{Gaeta2019Photonic-chip-basedCombs,Carlson2017Photonic-ChipFrequencies, Carlson2017Self-referencedWaveguides}. Optical reference cavities for laser stabilization have also been commercialized and portable systems have been developed \cite{Ohmae2021TransportableUncertainty,Origlia2018TowardsAtoms,Barbiero2022OpticallyCavity,Grotti2024Long-distanceClock,Zhang2020UltranarrowLaser, liu_36_2022}, however, the requirement to mitigate thermal noise has led to an increase in system complexity. A few of the primary optical-clock subsystems are the main targets for disruption, and the most important tasks are photonic integration of laser-beam delivery and simplification of laser-frequency and phase stabilization.

Integrated photonics encompasses various fields of research, including laser generation, manipulation, and detection \cite{Tran2022ExtendingPhotonics,Xiang20233DPhotonics,nader_heterogeneous_2025,jafari_heterogeneously_2024,Moody20222022Photonics}; integration of nonlinear optical processes \cite{Fathpour2018,Davanco2017HeterogeneousDevices,brodnik_monolithic_2025}; heterogeneous integration of material platforms \cite{Kaur2021HybridIntegration}; and nanophotonic devices like metasurface optics \cite{Arbabi2023AdvancesMetalenses}. For atomic-physics applications, several areas of integrated photonics have been developed to enhance system compactness and simplicity. 
Heterogeneously integrated III-V/SiN platform extends integrated photonics spectral coverage to sub-micrometer wavelength, beyond the bandgap of silicon \cite{Tran2022}.
Visible laser light can be delivered by integrated waveguides and emitted into free space by grating couplers, supporting integrated ion qubit systems and high fidelity operations, without inherently higher heating rates due to electric field noises from dielectric layer \cite{Mehta2016IntegratedQubit, Niffenegger2020IntegratedQubit, Ivory2021IntegratedIon}.  
A more recent demonstration showed the viability of a photonic integrated circuit (PIC) to route and emit 780 nm light as a 3D magneto-optical trap (MOT) for rubidium in a vapor cell \cite{Isichenko2023PhotonicTrap}. Still, challenges exist in the use of PICs, especially in three-dimensional MOTs, which require non-planar configuration of beams and  high optical power at visible wavelength. Various beamforming and low-loss waveguiding approaches have been explored \cite{Spektor2023UniversalPhotonics,Ropp2021Meta-gratingVisible, Ropp2023IntegratingChip,Yulaev2019Metasurface-IntegratedControl, Morin21, Chauhan22,Hsu2022Single-AtomTweezer}. 

Here, we explore a scalable infrastructure for Sr optical lattice clocks.  We leverage integrated photonics to stabilize the frequency of the cooling and trapping lasers, create the three-dimensional, multi-wavelength Sr MOT laser beam configuration without bulk optics, and we incorporate these elements into a compact package with ultrahigh vacuum, a Sr vapor source, and magnetic field coils. We describe two approaches towards versatile MOT beam control with a PIC and metasurface (MS) optics. A combination of PIC and MS optics simplifies the realization of MOT beams in three directions but suffers from high optical loss. With several MS optics integrated on a common fused-silica substrate, we realize a high efficiency, diverging-beam configuration for a MOT with high power, large beam size, and polarization control. MOT laser frequency stabilization is accomplished via frequency comb supercontinuum with fiber-packaged integrated nonlinear photonics. The waveguide supercontinuum devices enable use of a low-power comb system to stabilize the cooling narrow-linewidth lasers in a Sr lattice clock. Using the MS optics and supercontinuum devices, we demonstrate versatile diverging-beam MOTs of $^{84}$Sr, $^{86}$Sr, $^{87}$Sr, and $^{88}$Sr, with atom numbers reflecting their natural abundance that indicates precise design, tailoring, and robustness of the laser beam configuration. Our experiments demonstrate a set of scalable technologies for implementing Sr optical clocks, leveraging the design and integration capabilities of photonics.

\section{A scalable infrastructure for Sr magneto-optical trapping}

Operating an optical clock involves introduction of an atomic vapor from a solid, slowing atoms to the  velocity capture range of a MOT, trapping and cooling atoms in a MOT, and loading atoms into optical lattice for recoil-free spectroscopy \cite{Katori2003UltrastableTrap}. Iterative spectroscopy of the ultra-narrow atomic clock transition with a pre-stabilized laser enables constant tracking of the resonance. Complex maintenance and data analysis operations are also required to stabilize the clock laser frequency \cite{Collaboration2021FrequencyNetwork}.
For a Sr optical lattice clock, $^{87}$Sr is of special interest as its $^{1}S_0 \leftrightarrow$ $^{3}P_0$ weakly allowed transition provides one of the highest performance references for a clock \cite{Aeppli2024}, while $^{88}$Sr is also convenient due to its simpler atomic structure and higher natural atomic abundance \cite{Xu2003CoolingStrontium, barbiero_sideband-enhanced_2020}. The use of Sr necessitates integration of laser beam wavelength ranging from 461 nm to 813 nm, posing a challenge in the design and fabrication of ultralow-loss integrated photonics.

Figure 1a presents the conceptual design of our integrated photonics Sr optical-clock system, including the atomic Sr beam generated in ultrahigh vacuum (UHV), Doppler slowing of the Sr beam, and the three-dimensional, two-color nanophotonics MOT system \cite{jammi_three-dimensional_2024,ferdinand_laser-cooling_2025}. We design for the MOT beams to be emitted from planar substrates consistent with integrated photonics, therefore we create a flattened geometry for the UHV chamber and offset-planar anti-Helmholtz magnetic coils. In a MOT, there is a tradeoff between laser beam intensity and volume of the trap \cite{Lindquist1992ExperimentalTrap}. Whereas with integrated photonics there are substantial tradeoffs between beam size and fabrication complexity, specifically the size of a beam emitted from a grating coupler due to numerical aperture and the fabrication size of metasurface optics for millimeter or centimeter size laser beams. Therefore, in our system design, we use a diverging beam MOT geometry \cite{nichols_magneto-optical_2020}, and we avoid Zeeman slower or 2D MOT due to their inherent large volume; a single Doppler slowing beam is used instead to boost trapping efficiency \cite{LudlowPRL2018}.

To assemble the vacuum system, we integrate a commercial effusive oven with a microcapillary-array collimation nozzle into a standard UHV apparatus. The atom source is coupled to a titanium chamber with large-diameter welded and brazed windows that provide multi-axis optical access. These windows are anti-reflection coated for angled incidence at MOT wavelengths. An appendage along the Sr atomic-beam path houses a sapphire window for the slowing beam and is actively heated to prevent Sr deposition. Through this window, a 461-nm laser beam, detuned from the \(^{1}S_0\!-\!^{1}P_1\) transition and counter-propagating to the atomic beam, selectively slows a velocity class and enhances capture efficiency in the three-dimensional MOT. A switchable mechanical shutter in the atomic-beam path blocks both atoms and thermal radiation from the oven when required, while a narrow-diameter tube upstream of the science chamber provides differential pumping and thermal isolation from the hot oven. During operation, the oven is heated to 673~K and the science-chamber pressure is maintained at \(10^{-9}\)~Torr. This single-beam slowing approach~\cite{LudlowPRL2018} provides a simpler alternative to Zeeman slowers and two-dimensional MOTs, which can deliver higher and colder atom fluxes at the cost of increased system complexity.

\begin{figure*}[t] \centering
\includegraphics[width=0.95\linewidth]{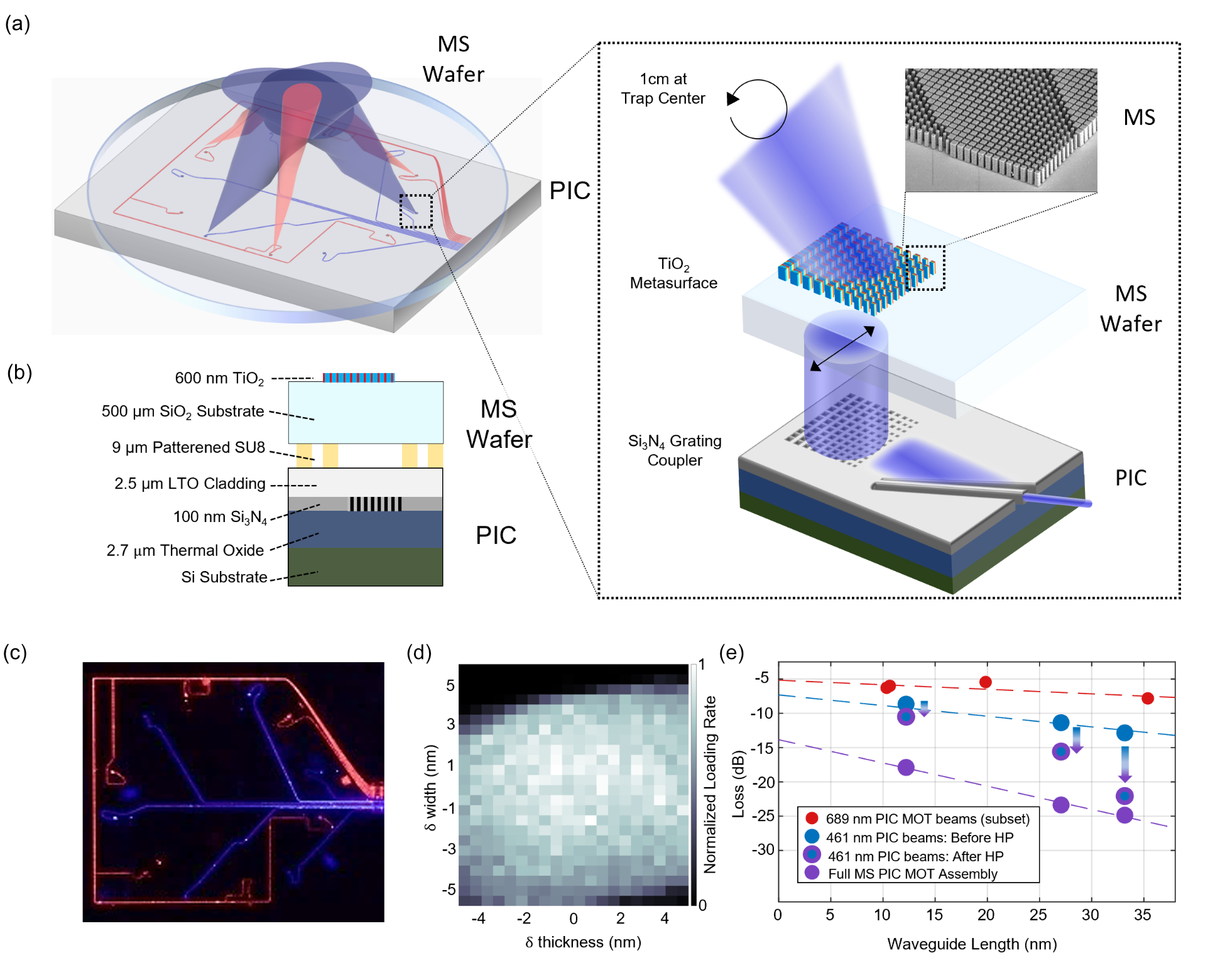}
\caption{\textbf{PIC–MS approach for MOT beam generation.}
(a) Waveguides on a PIC chip route \SI{461}{nm} and \SI{689}{nm} light to apodized meta-grating couplers, which vertically outcouple light to a wafer with MS optics for Sr MOT beam formation.
(b) Material layer stack of the PIC–MS architecture.
(c) Photograph of a fabricated PIC routing \SI{461}{nm} (broad-line) and \SI{689}{nm} (narrow-line) strontium MOT light.
(d) Monte Carlo simulations of relative MOT loading rates versus fabrication imperfections in waveguide width and thickness.
(e) Loss characterization of the PIC–MS system. Low-power propagation loss is measured with on-chip powers of $\sim$1 mW at \SI{461}{nm} and \SI{689}{nm} (red and blue circles). After illumination at high power, length-dependent loss increases significantly for the PIC (purple–blue) and the full PIC–MS system (purple), with optical powers of 50 mW and 150 mW, respectively.} \label{fig:fig2}
\end{figure*}

To generate laser beams and the magnetic field gradient for the MOT, we create a system of MS optics on fused-silica wafers and displaced coils above and below chamber; see Fig. 1b. Each MS imparts an engineered phase profile to the wavefront of the incident light and allows nearly arbitrary control of beam properties, including angle, divergence, and polarization \cite{jammi_three-dimensional_2024}. In the experiments presented here, we implement phase profiles to deflect input beams by 45 degrees to propagate to trap center, diverge the output beam to achieve desired optical intensity at MOT position, and rotate linear polarization of input light to circular polarization required for atom trapping. In our setup, we accommodate two flat anti-Helmholtz coils parallel to the plane of metasurface substrates, while spatially offset them in axial direction for best optical access, as shown in Fig. \ref{fig:fig1}b. One pair of the MOT beams propagates through the coils, and we fix the three identical MSs by aligning wafer flats to a custom mounting structure. The corresponding polarizations of each MOT beam are controlled by illuminating MSs with polarization-maintaining fiber outputs in designed angles to the MSs.

For the overall system design of the planar metasurface optics and magnetic field gradient, we perform Monte-Carlo simulations of laser-cooled atomic trajectories with this geometric design to predict its ability to trap atoms \cite{barbiero_sideband-enhanced_2020}. We simulate MOT loading rates with different design configurations to optimize the system, including MOT beam geometry, gradient coil locations, Doppler slowing beam, and effusive oven. Figure \ref{fig:fig1}c shows simulated atomic trajectories for our finalized system design, with parameters commonly used in our real experiments: total \SI{461}{nm} MOT beam intensity $I_{\rm{MOT}}$ = \SI{40}{mW\per cm\squared}, frequency detuning $\Delta_{\rm{MOT}} \approx - 1.5~\Gamma $ with $\Gamma = 2 \pi\; \times$ \SI{31}{MHz} being the linewidth of the transition, magnetic field gradient \SI{50}{G\per cm}, and slowing beam power $P_{\rm{slowing}}$ = \SI{2}{mW} with a \SI{5}{mm} waist and detuning $\Delta_{\rm{slowing}}\approx - 5.5~\Gamma$.

With our experimental setup, we demonstrate laser cooling and trapping of up to \SI{4E+5}{} $^{87}$Sr atoms, as pictured in Fig. \ref{fig:fig1}d. For this experiment, we assemble the fused silica wafers with the metasurface optics and holders for the bare optical fiber that deliver 461 nm trapping light into two precision-machined aluminum mounts, and we assemble those elements around the vacuum system. Upon coupling light into the fibers, we detect the MOT without any further alignment of the system. Here, the effusive oven is operated at \SI{400}{\celsius}, and more atoms could be trapped with higher oven temperature at the expense of vacuum pressure. 

\section{Integrated photonics for MOT beam generation}

We explore two approaches towards the generation of MOT beams with integrated photonics: (1) a PIC system composed of both in-plane waveguide routing and out-of-plane emission of optical beams, and (2) a metasurface optics system integrated on a fused-silica wafer with illumination by bare, polarization-maintaining optical fibers. We characterize the performance of both approaches at 461 nm and 689 nm, which are the target wavelengths in Sr to operate broad and narrow-linewidth laser cooling, respectively. Our experiments focus on evaluating the efficiency and performance of the integrated photonics systems with respect to requirements for practical MOT operation. A challenge in the use of PICs for MOT beam emission is creation of circular polarization; though, grating couplers can be adapted to emit circular polarization by interference of two waveguide modes \cite{Spektor2023UniversalPhotonics}. In our MOT experiments, we adopt the metasurface-optics approach due to workable conversion efficiency of laser cooling light from optical fiber. 

\subsection{Hybrid PIC and MS optics approach}

In the PIC-MS approach (Fig. 2a), the PIC provides routing and beam emission for 461 nm and 689 nm. At each grating coupler, light from a waveguide couples into a collimated slab-mode beam through an engineered tapered gap between the waveguide and the slab \cite{Ropp21}. For emission, the collimated slab-mode propagates to an apodized meta-grating coupler, which couples the light out of the plane of the device to create a free-space Gaussian beam incident on a metasurface optic \cite{Ropp2023IntegratingChip}. The metasurface optic deflects the beam, controls the divergence, and creates the circular polarization for operation of the MOT. We fabricate all six of the metasurface optics on a fused-silica wafer, which is die bonded to the PIC with 10~$\mu$m precision by use of a patterned SU8 layer. We show the materials stackup for this approach in Fig.~\ref{fig:fig2}~b. This hybrid PIC and metasurface integration approach offers a versatile palette for MOT beam generation. 

The PIC is composed of a silicon substrate with a single device layer of stoichiometric silicon nitride (SiN) deposited via LPCVD. The SiN device layer thickness is 100 nm, the lower cladding of thermally oxidized silicon is 2700 nm thick, and the upper cladding is 2500 nm of low-temperature SiO$_2$ (LTO) deposited by LPCVD. We fabricate the PIC on a 100 mm silicon wafer with electron-beam lithography (EBL), reactive ion etching, the LPCVD processes, and thermal annealing. Laser light is coupled to the PIC from lensed optical fiber in an array via edge couplers in the SiN layer; for independent control of optical power we use separate fibers for each MOT beam. The waveguides that compose the PIC feature six branches that deliver light to apodized meta-grating couplers, which nearly vertically transfer light out of the PIC. The meta-gratings are located along a circle of 25.4 mm diameter, which is designed in accordance with the overall system arrangement of the UHV chamber and the magnetic field coils. At each emission grating, light from a waveguide couples into a collimated slab-mode beam through an engineered tapered gap between the waveguide and the slab \cite{Ropp21}; see the inset of Fig. 2a. The nearly vertical beam emitted by the grating is linearly polarized and has a Gaussian intensity profile.

\begin{figure*}[t]
\centering
\includegraphics[width=0.98\linewidth]{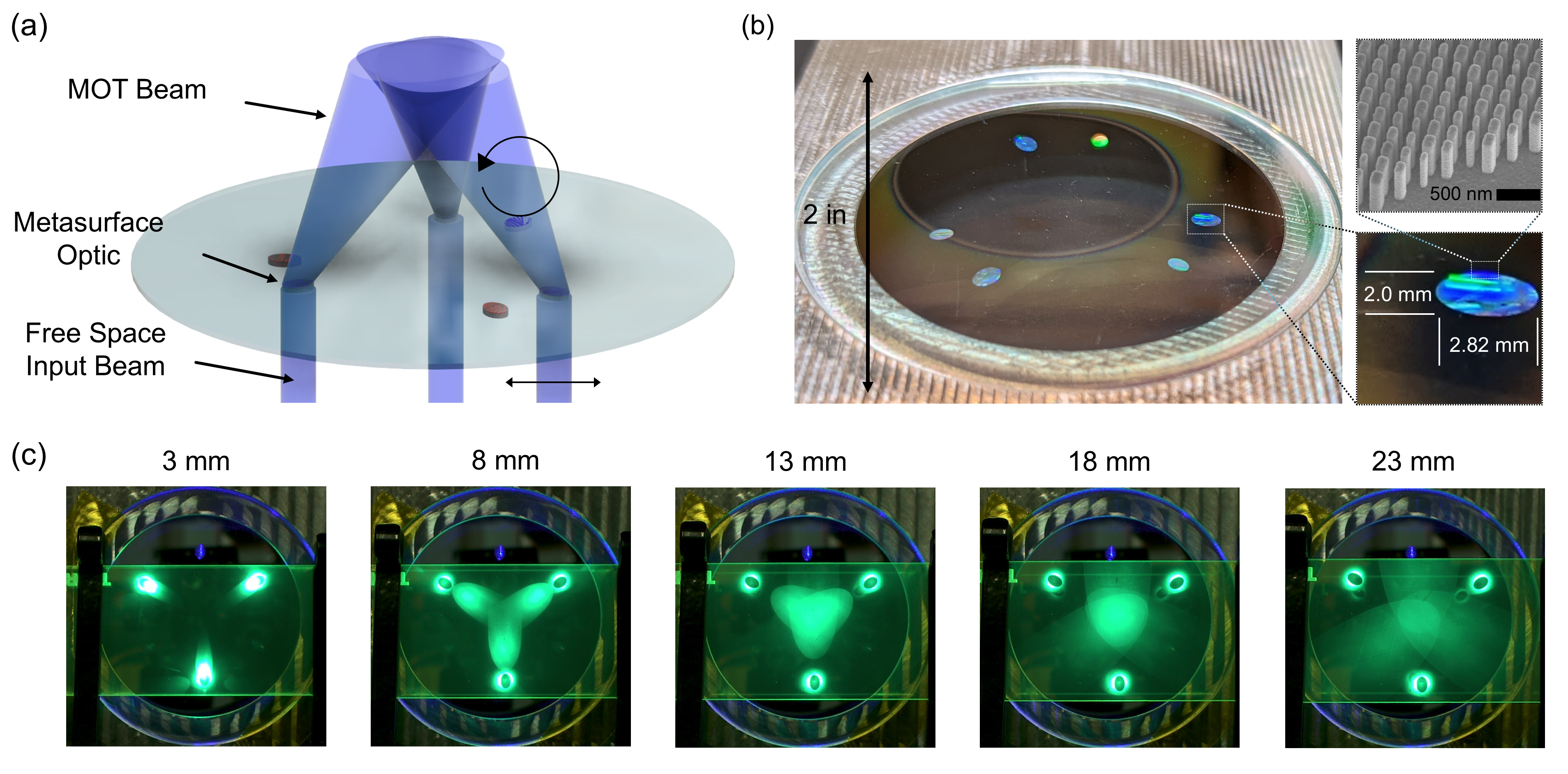}
\caption[\parbox{6.5in}]{ \textbf{MS optics approach for MOT beam generation} \textbf{(a)} Illustration of MOT beam generation with direct illumination of multifunctional metasurface optics fabricated on a common substrate with free space beams. \textbf{(b)} Photo of \SI{461}{nm} and \SI{689}{nm} MS optics fabricated on common substrate. Subpanels show elliptical dimensions of a 461 nm metasurface optics and an SEM image of constituent nanopillars \textbf{(c)} Propagation, expansion, and intersection of MS derived \SI{461}{nm} MOT beams viewed on a fluorescent plate at the plate is translated vertically above the MS wafer.}
\label{fig:fig3}
\end{figure*}

For the hybrid PIC and metasurface approach, we fabricate metasurface optics on a fused silica substrate with titania (TiO$_2$) nanopillars. Using a thick layer of electron-beam resist, we use EBL to pattern an array of nanopillars. The titania nanopillars are grown via high-rate atomic layer deposition into the EBL resist, and subsequent liftoff of the resist creates the designed pattern of the metasurface optic. The full transformation to MOT trapping beams is accomplished by the titania metasurface optics, which deflect the beams toward the trap center, control the beam divergence, convert to the appropriate circular polarization. In the designed system, the upward beams are transformed by transmissive MS optics and the downward beams are formed by reflective MSs on a second wafer not shown in Fig. \ref{fig:fig2}.  The two chips are bonded together with a patterned SU8 layer between the MS SiO$_2$ substrate and PIC LTO cladding; see Fig. \ref{fig:fig2}(b) for full layer stack.  The gratings and MSs are laid out on \SI{25.4}{mm} diameter circle.  Figure \ref{fig:fig2}(C) shows a fabricated PIC chip that has been illuminated with 461 nm and 689 nm light.

A view of the promise and challenge of PICs for beam routing and emission in a Sr optical clock and similar quantum sensing and information systems is seen in the device realization and characterization data presented in Fig.~\ref{fig:fig2}(c) and (d). We realize a PIC routing and emitting light for all twelve MOT beams of two wavelengths for both the blue MOT and the subsequent the first broad-line cooling stage of Sr and subsequent narrow-line cooling needed to load atoms into an optical lattice;  Fig.~\ref{fig:fig2}~c. This highlights the potential for PICs to disrupt traditional laser cooling technologies by simplifying generation of complex beam geometries. One concern in realizing an atom-photonic interface with a PIC and MS based approach are imperfections and uncertainties in the PIC fabrication, which lead to pointing errors, divergence errors, and displacements in our MOT beams.  These fabrication errors lead to distorted trapping beams and usually a reduced trapping volume.   We quantify the influence of these errors on with Monte Carlo simulations by monitoring MOT loading rate changes resulting from imperfections in PIC feature height and width up to $\pm$\SI{5}{nm} and $\pm$\SI{6}{nm}, respectively in \SI{0.25}{nm} increments, as shown in Fig.~\ref{fig:fig2}~d. We calculate the loading rate from 7400 atom trajectories for each set of fabrication imperfection values. Our simulations show robust trapping up to  $\pm$\SI{5}{nm} in feature thickness and $\pm$\SI{4}{nm} in feature width, within the capabilities of high precision nanofabrication, and greater than $80\%$ of peak trapping efficiency within $\pm$2 nm.

Despite the benefits of the PIC-MS architecture for direct integration and ultra-compact systems, this beam delivery approach suffers from high optical propagation loss in the routing layer, especially in the blue portion of the visible. Given the moderate power available from conventional diode laser sources, this represents an obstacle for realizing a MOT. We investigate losses in the PIC chip at both 461 nm and 689 nm wavelengths under three different experimental conditions, the results of which are presented in Fig.~\ref{fig:fig2}~e.  Since each waveguide has a different length, we can distinguish between coupling losses and optical propagation losses by analyzing the loss as a function of propagation length.  At $\sim$ mW level optical powers in the PIC, we measure  \SI{0.66}{dB\per cm}  propagation loss and \SI{5.2}{dB} residual coupling loss at \SI{689}{nm} (red circles with red linear fit) and \SI{1.6}{dB\per cm}  propagation loss and \SI{7.3}{dB} coupling loss at \SI{461}{nm} (blue circles and blue linear fit). Loss measurements in the complete PIC-MS atom photonic interface at higher powers  required for trapping (50 mW) showed total coupling losses associated with fiber-waveguide and grating-MS transfer (\SI{13.8}{dB} total) as well as increased length dependent coupling losses of  \SI{3.4}{dB\per cm} (purple circles and purple fit). The 461 nm waveguides that were not exposed to high powers did not show higher losses.  The correlation between increased losses and waveguide length seen in the complete system (purple) and in PIC measurements after high power exposure (blue and purple circles) suggests power-dependent damage.  While this platform offers promise for a fully integrated atom-photonic interface, the optical losses susceptibility to photo-degradation at the 461 nm powers required for trapping make use in our current experiments impractical. 

\begin{figure*}[t] \centering
\includegraphics[width=0.95\linewidth]{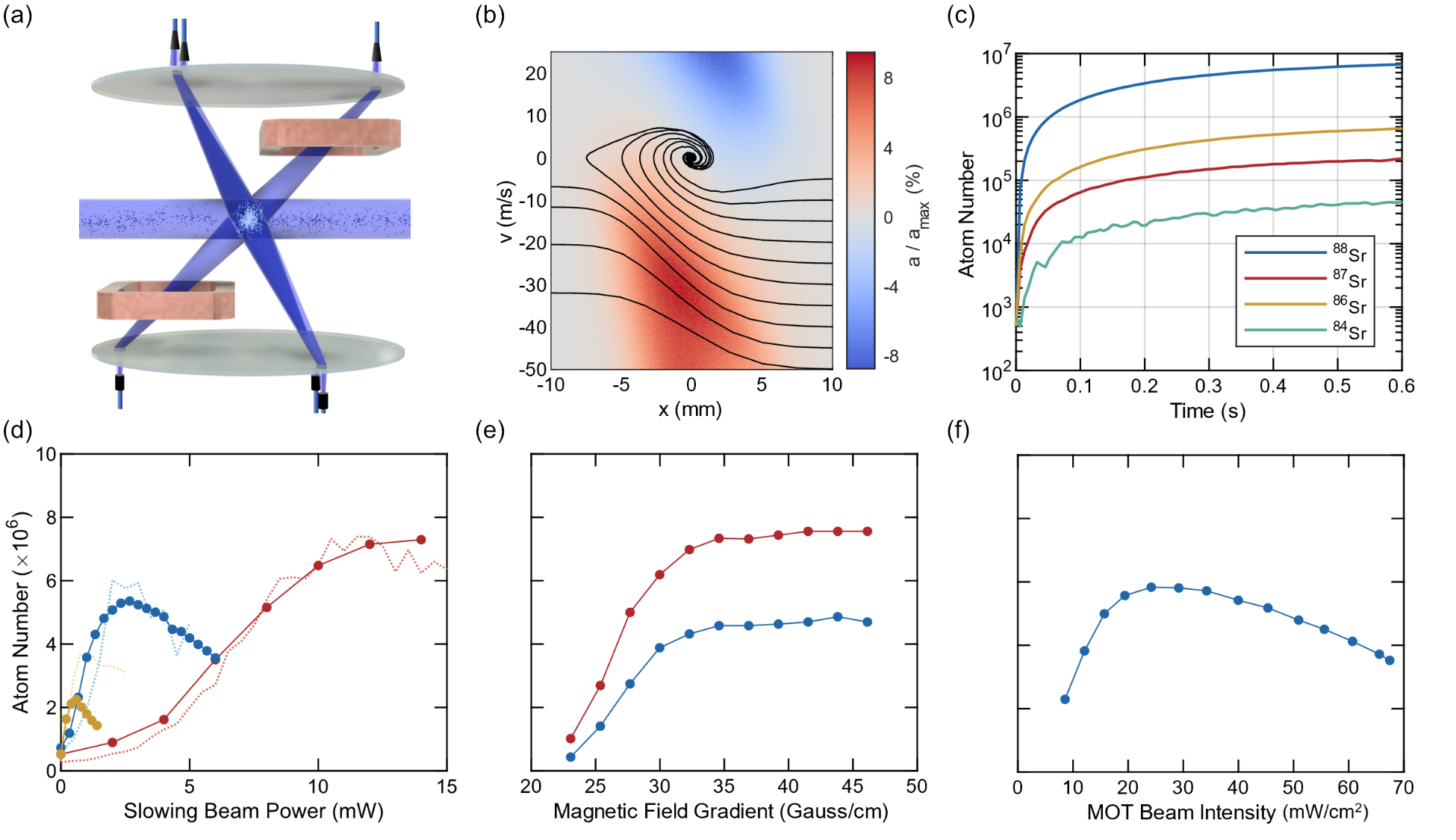}%
    \caption{ \textbf{Characterization of Sr MOT with MS optics. (a)} Illustration of the MOT geometry with MS optics and the slowing laser beam that assists MOT loading from the thermal atom beam. 
    \textbf{(b)} Monte Carlo simulations of laser-cooled atomic phase-space trajectories for MS MOT geometry depicted in panel (a). 
    \textbf{(c)} Experimental MOT loading curves for each naturally occurring stable isotope of strontium. Experimental parameters include: magnetic field gradient, $\nabla |\textbf{B}|$ = \SI{39}{G\per cm}, total (six-beam) intensity $I_{\rm{MOT}}$ = \SI{40}{mW\per cm\squared}, $P_{\rm{slow}}$ = \SI{2.6}{mW}, and $\Delta_{\rm{slow}} = -5.5~\Gamma$. 
    \textbf{(d)} $^{88}$Sr MOT atom number as a function of slowing beam power for three detunings: $\Delta_{\rm{slow}} = -3.5~\Gamma$ (gold), $\Delta_{\rm{slow}} = -5.5~\Gamma$ (blue) and $\Delta_{\rm{slow}} = -7~\Gamma$ (red); $\nabla |\textbf{B}|$ and $I_{\rm{MOT}}$ are the same as in panel (b). Monte Carlo simulations plotted as dotted lines show agreement with the experimental data. 
    \textbf{(e)} $^{88}$Sr MOT atom number as a function of axial magnetic field gradient with $I_{\rm{MOT}}$ = \SI{46.5}{mW\per cm\squared} for two different conditions of the slowing beam: $\Delta_{\rm{slow}} =- 7~\Gamma$ with $P_{\rm{slow}}$ = \SI{14}{mW} (orange), and $\Delta_{\rm{slow}} = -5.5~\Gamma$ with $P_{\rm{slow}}$ = \SI{2.0}{mW} (blue). 
    \textbf{(f)} $^{88}$Sr MOT atom number as a function of $I_{\rm{MOT}}$ and slowing beam parameters $\Delta_{\rm{slow}} = -5.5~\Gamma$ and $P_{\rm{slow}}$ = \SI{2.0}{mW}. }
\label{fig:fig4}
\end{figure*}

\subsection{MS optics approach}
\label{MS_sec}

We also investigate the performance of a MS-only system that is illuminated with optical beams shaped from emission and routed from the source laser by low-loss, polarization-maintaining optical fiber; see Fig. \ref{fig:fig3}. This system benefits from fundamental MS efficiency of up to $\approx 50 \%$ and a significantly increased damage threshold due to higher mode area and much lower intensity. These two characteristics allow us to generate several MOT beams with at least several mW of power, based on a low-noise, single frequency 461 nm laser. Figure \ref{fig:fig3}(a) shows a photo of a MS wafer with an illustration of how the input free space beam is transformed by the MS to form a MOT beam. The wafer contains three 461 nm MSs for broad line cooling and three 689 nm MSs for narrow-line cooling; see Fig. \ref{fig:fig3}(b).  Both the 461 nm and 689 nm MSs are composed of equal height TiO$_2$ nanopillars (upper inset of Fig. \ref{fig:fig3}(b)) is designed to steer the beams towards the trap center, alter the beam divergence and convert the input linear polarization to the appropriate handedness of circular polarization \cite{jammi_three-dimensional_2024}.  Atom trapping experiments at 461 nm are described later in this work and trapping experiments at 689 nm using a fully fiber coupled system with two MS optics is described in Ref. \cite{Ferdinand2024LaserSystem}. We fabricate the metasurface optics in an elliptical shape  such that the output MOT beam is circularly symmetric in the plane normal to the beam propagation direction; see lower inset of Fig. \ref{fig:fig3}(c). Figure \ref{fig:fig3} (c) shows representative MOT beam propagation as a function of height above the wafer with all beams converging on the trap center. 

\section{Characterization of the Sr MOT with MS optics}

With the MS optics approach that generates the complex laser-beam configuration required for a MOT, we explore trapping different Sr isotopes on the 461 nm transition; see Fig. \ref{fig:fig4}. This provides a test of the precision with which MS optics can realize a MOT beam configuration. In this approach, illustrated in Fig. \ref{fig:fig4}(b), arrays of multifunctional MS elements fabricated on two fused-silica substrates are illuminated with commercial polarization-maintaining (PM) fiber \cite{jammi_three-dimensional_2024}. Both substrates and all PM fiber ferrules are mechanically fixed in a rigid custom mount, producing the full set of MOT beams without bulk optics or beam alignment. Atoms from the effusive source, however, are too fast for direct capture because the Doppler shift pushes them out of resonance with the MOT beams. A counter-propagating free-space slowing beam is therefore introduced to enable effective trapping. Monte Carlo simulations of the trapping dynamics (Fig. \ref{fig:fig4}a) show a capture velocity of $\sim$30 m/s for $I_{\rm MOT}=1.2\times I_{\rm sat}$, consistent with phase-space trajectories in Fig. \ref{fig:fig4}(b). Adding a slowing beam with power $P_{\rm slow}=5$ mW and detuning $\Delta_{\rm slow}=-5 \Gamma$ extends the capture velocity along the beam axis to $\sim$130 m/s, substantially increasing the fraction of atoms trapped in the MOT.

\begin{figure*}[ht!]
\centering
\includegraphics[width=0.95\linewidth]{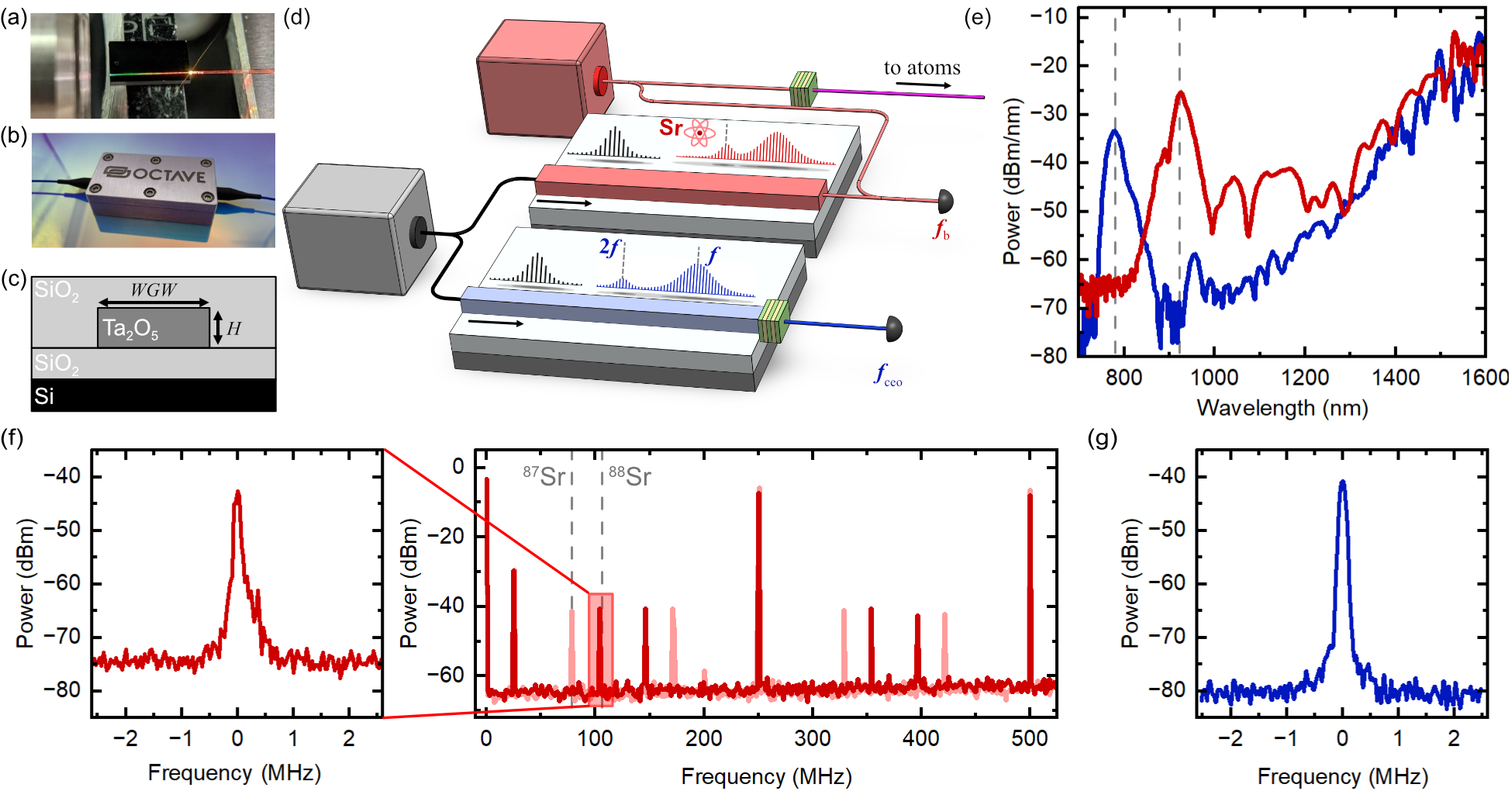}
\caption[\parbox{6.5in}]{ \textbf{Photonic chip supercontinuum (SC) for laser frequency stabilization.} (a) A photo of a visible supercontinuum generation in nonlinear TaO waveguides. (b) 3D mechanical drawing a of a packaged SC chip with integrated fiber pigtails. (c) Tantala photonics platform layer stack (d) Laser system and SC modules with dispersive wave (DW) emission at 922 nm for stabilization of the Sr cooling lasers (top, red) and at 780 nm for $f$-$2f$ stabilization of $f_{\text{ceo}}$ (bottom, blue). Second harmonic generation from 922 $\rightarrow$ 461 nm generates the appropriate wavelength for broad line Sr cooling. (e) Optical SC spectra of the Sr SC module (red) and $f_{\text{ceo}}$ module (blue); Dashed lines indicate DW design wavelength of 922 nm and 780 nm respectively. (f) Heterodyne beats ($f_{\text{b}}$) between Sr module SC and our trapping laser at 922 nm.  (f,right panel) Heterodyne beats when the 922 nm laser is tuned to the frequency to neccessary for trapping $^{88}$Sr (dark red) and $^{87}$Sr (light red).  The respective LO oscillator frequencies for trapping $^{88}$Sr and $^{87}$Sr are indicated with vertical grid lines. (f,left panel) Enlargement of the unlocked heterodyne beat used to trap $^{88}$Sr (32 dB SNR at RBW = 100 kHz).  (g) The unlocked $f_{\text{ceo}}$ beat, detected at 780 nm (37 dB SNR at RBW = 100 kHz).  
}
\label{fig:fig5}
\end{figure*}

We realize MOTs of all naturally occurring stable isotopes of strontium with the MS optics approach, as shown in the loading curves of Fig. \ref{fig:fig4}(c). The relative MOT atom numbers follow the natural isotopic abundances. Trapping a specific isotope is achieved by tuning the repump and trapping laser frequencies, with details of our stabilization method given in Fig. \ref{fig:fig5}. Two repump lasers at \SI{707}{nm} and \SI{679}{nm} enhance cooling in the \SI{461}{nm} MOT. Efficient trapping of fermionic $^{87}$Sr additionally requires scanning the repump lasers across the hyperfine manifold over a few GHz, which we implement by directly ramping the piezoelectric transducers inside the laser cavities. The trapping efficiency for $^{87}$Sr could be further improved with more sophisticated hyperfine addressing schemes. The successful trapping of $^{87}$Sr demonstrates the capability of our compact MOT system to confine atoms with complex internal structure.

In Fig. \ref{fig:fig4} (d,e,f) we present experimental measurements and simulations of the MOT, utilizing the counter propagating beam geometry of Fig. \ref{fig:fig3}(a). The MS optics setup mirrors that of the PIC MS architecture, fabricated on a 25.4 mm circle to create three circularly polarized MOT beams with 1 cm diameter at the trap center. We replace the PIC and illuminate the metasurface optics directly with millimeter diameter beams. To optimize MOT performance, we first explore the dependence on slowing beam parameters $P_{\rm{slow}}$ and $\Delta_{\rm{slow}}$; see Fig. \ref{fig:fig4}(d). We measure atom number in the MOT as a function of slowing beam power with three settings of slowing beam detuning. Total MOT beam intensity $I_{\rm{MOT}}$ is 46 mW/cm$^2$ and magnetic field gradient is \SI{39}{G/cm} for these experiments. The maximum atom number and corresponding $P_{\rm{slow}}$ increase with detuning. Our Monte Carlo simulations (dotted lines) reproduce these trends, with results scaled by two global parameters to account for slowing-beam power attenuation and MOT atom loss. In Fig. \ref{fig:fig4} (e), we build on the the results of Fig. \ref{fig:fig4}(d) and examine how $\nabla|\textbf{B}|$ affects atom number for two slowing beam configurations based on previous optimization results. In both cases the atom number increase with increasing gradient up to \SI{35}{G/cm}, where the atom number starts to plateau.  
In Fig. \ref{fig:fig4}(f) we measure the atom number of $^{88}$Sr as a function of total peak intensity of the MOT beams, showing a maximum of \SI{6E+6}{} at an intensity of \SI{30}{mW\per cm\squared}. This experimental characterization of the unique MS optics MOT geometry supports system design and optimization for optical lattice clocks and other applications of ultracold alkaline earth samples.

\section{Laser frequency stabilization with integrated photonics supercontinuum}

In this work, we frequency-stabilize the lasers by use of frequency-comb supercontinuum generated in integrated photonic devices \cite{Lamee2020NanophotonicNm}. Figure \ref{fig:fig5} illustrates our approach to wavelength referencing across the visible and near-IR spectrum required for Sr optical lattice clock operation. Our method employs photonic-chip supercontinuum from narrowband frequency combs to generate the near-IR and visible light needed for addressing Sr transitions and for $f$--$2f$ self-referencing. The chips are based on our tantala integrated photonics platform \cite{Jung2021TantalaPhotonics} and advances in dispersion engineering of nonlinear optical waveguides \cite{Black2021Group-velocity-dispersionPhotonics}. Figure \ref{fig:fig5}(a) shows a laboratory test of an air-clad tantala chip producing visible supercontinuum, while Fig. \ref{fig:fig5}(b) shows packaged, fully oxide-clad devices integrated into compact enclosures with PM-fiber pigtails for robust coupling in laser-stabilization experiments. With a film thickness of 800 nm, our tantala waveguides support broadband supercontinuum from the UV to the mid-IR. Control of the waveguide width (WGW) tunes the dispersion and sets the wavelengths of dispersive waves (DWs), which enhance power of the supercontinuum at the specific   wavelengths need for a Sr lattice clock.  

The two modules used in this work are shown schematically in Fig. \ref{fig:fig5}(d); the top oxide is omitted for clarity. Both are pumped by a 1550 nm erbium fiber modelocked laser system with $f_{\text{rep}} \approx 250$ MHz and pulse energy $\sim$150 pJ. For absolute referencing of the Sr trapping laser (top panel), we use $WGW = 2.15~\mu$m to generate a DW at 922 nm; subsequent frequency doubling produces 461 nm light for the broad-line MOT of Fig. \ref{fig:fig4}. For comb $f_{\text{ceo}}$ stabilization (bottom panel), we use $WGW = 1.4~\mu$m to generate a DW at 780 nm. This module incorporates a short section of periodically poled lithium niobate (PPLN) at the output to generate 780 nm from 1560 nm ($f \rightarrow 2f$). Heterodyne detection between the 780 nm second harmonic and the 780 nm DW yields an $f_{\text{ceo}}$ beat. Figure \ref{fig:fig5}(e) shows spectra from the 922 nm (red) and 780 nm (blue) modules, with vertical dashed lines marking the designed DWs. We generate --26 dBm/nm at 922 nm (1.8 nW per comb tooth) and --34 dBm/nm at 780 nm (0.2 nW per comb tooth).  

Figure \ref{fig:fig5}(f) presents the optical heterodyne beats between our 922 nm CW laser and the 922 nm supercontinuum output. The left panel shows the beat on a 5 MHz span with 32 dB SNR in 100 kHz resolution bandwidth (RBW). The right panel shows the photodetector output when the 922 nm laser is tuned to trap $^{88}$Sr (dark red) and $^{87}$Sr, demonstrating simple frequency tuning between isotopes. For each detuning, the RF spectrum contains the first and second harmonics of $f_{\text{rep}}$ along with two heterodyne beats, $f_{\text{b1}}$ and $f_{\text{b2}}$, located between consecutive $f_{\text{rep}}$ harmonics. These beats correspond to heterodyne between comb teeth $n$ and $n+1$. Locking one beat to a local oscillator (LO) at $f_{\text{LO}}$ fixes the CW laser frequency as $\nu_{922} = n f_{\text{rep}} + f_{\text{ceo}} + f_{\text{LO}}$. The appropriate LO frequencies for trapping $^{88}$Sr and $^{87}$Sr are indicated in Fig. \ref{fig:fig4}(c), and their difference corresponds directly to the isotope shift $\Delta \nu_{\text{iso}}$ via $f_{\text{LO}}^{(88)} - f_{\text{LO}}^{(87)} = \Delta \nu_{\text{iso}}/2$. Finally, Fig. \ref{fig:fig5}(g) shows the $f_{\text{ceo}}$ beat at 780 nm with 40 dB SNR in 100 kHz RBW.

\section{Compact, integrated system for ultracold Sr}

\begin{figure*}[t] \centering
\includegraphics[width = 0.95\linewidth]{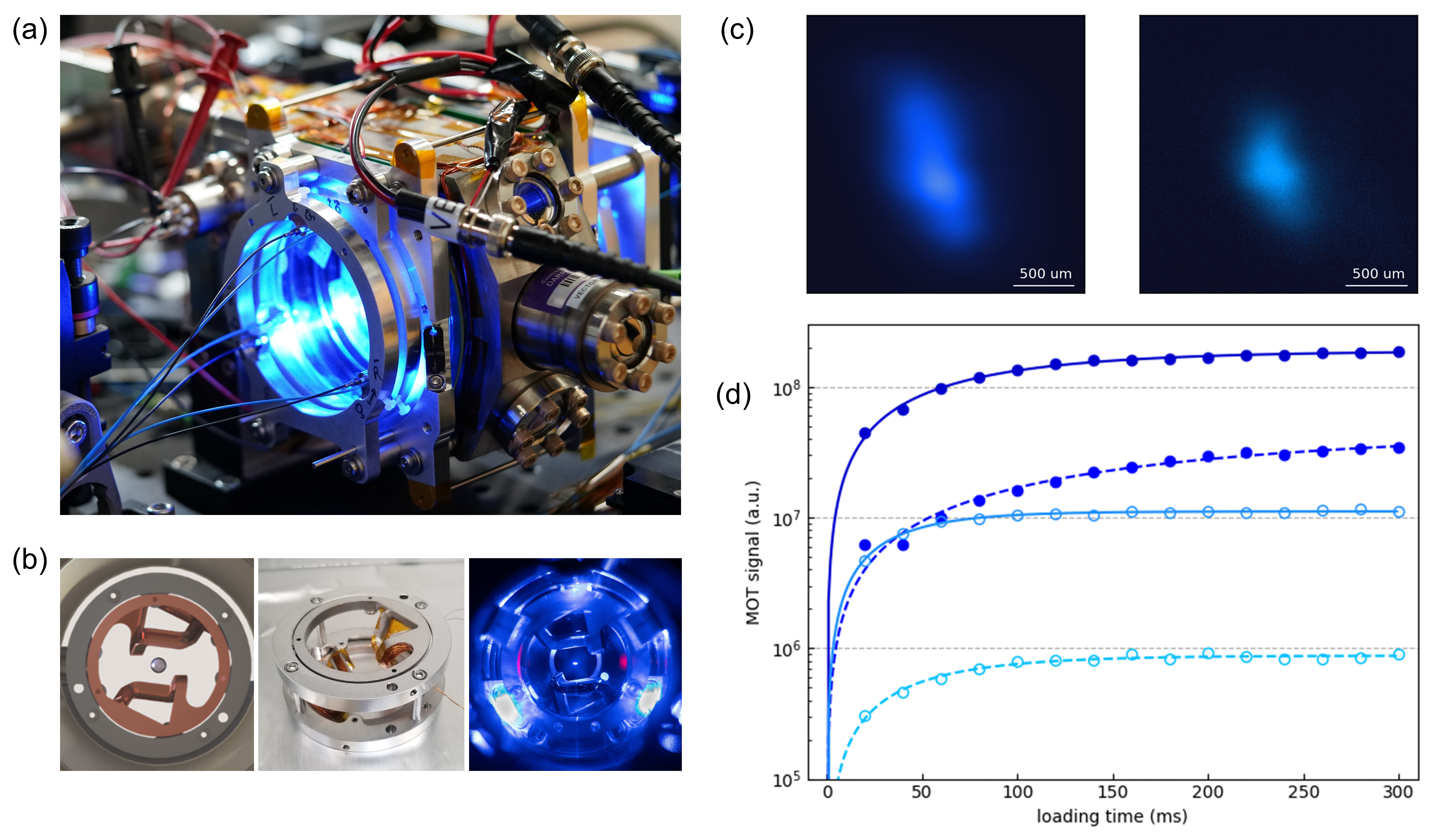}
\caption{\label{fig:fig6} \textbf{MS MOT demonstration with engineering development system. (a)} Photo compact vacuum chamber with laser controlled source, in-vacuum coils and integrated fiber-coupled MS beam delivery system during broad-line MOT operation.  Trapping laser sources and imaging system is not visible in this image.  \textbf{(b)} UHV compatible, in-vacuum anti-Helmholtz coils. Left: Drawing of the coil mounts. 
 Middle: Finished coil assembly before installation. Right: Fully operational in-vacuum coils during a \SI{461}{nm} MOT sequence; atom fluorescence is visible in the center of the image.  \textbf{(c)} Fluorescence images of trapped $^{88}$Sr (\SI{E+6}{} atoms, dark blue, left) and $^{87}$Sr (\SI{E+5}{} atoms, light blue, right) trapped with a broad MOT. The trapping beams are also used for imaging with identical $I_{\rm{MOT}}$ and $\Delta$.  \textbf{(d)} MOT loading curves with fits for $^{88}$Sr (closed circles) and $^{87}$Sr (open circles) at chamber pressures of \SI{3E-8}{Torr} (solid lines) and \SI{1E-9}{Torr} (dashed lines).}
\end{figure*}

To explore system integration with our scalable infrastructure, we have performed a full cycle of engineering development to create one physics package that contains a UHV chamber, a laser-heated Sr vapor source, magnetic field coils for the MOT gradient and biasing, and the metasurface optics system to create the blue MOT laser beams. The volume of the system is roughly 0.5 L, accounting for the primary components and excluding the external camera imaging beam path. Figure \ref{fig:fig6}(a) shows the package during blue MOT trapping operation, and using the package we have demonstrated blue MOT trapping of all the stable Sr isotopes ($^{84}$Sr, $^{86}$Sr, $^{87}$Sr, $^{88}$Sr). Further details of the optimized MS beam delivery systems are described in Refs. \cite{jammi_three-dimensional_2024, ferdinand_laser-cooling_2025}. 

The vacuum chamber is designed to be compact while allowing sufficient optical access for various laser beams through two large windows. We install a customized mounting structure for MS wafers onto the vacuum chamber, eliminating the alignment of trapping beam intersection to the chamber center and magnetic field coils. An anti-Helmholtz coil pair is installed inside the vacuum chamber to provide the MOT gradient with reduced power consumption and cooling requirements. The coils use a twisted geometry to maximize the optical access to MOT trapping beams while operated in close proximity to the trapping region (Fig. \ref{fig:fig6}b). We typically run \SI{1}{\ampere} of current during the \SI{461}{nm} MOT stage for a magnetic field gradient of \SI{50}{G/cm} in the direction of the MOT beams.

To generate a vapor of Sr inside the vacuum chamber, we use a CW diode laser to heat a Sr pellet. This atomic vapor source features a low total electrical power consumption of $\sim$2 W and $<$ 5 minute warm-up time compared to the conventional heated capillary-oven source (Fig. \ref{fig:fig1}). The Sr pellet is placed within a few cm distance from the intersection region of trapping beams, enabling efficient MOT trapping without a Zeeman slower or other measures. The small heated area of the source also reduces overall gas load of the system, making it compatible with a compact 3 L/s ion pump for a vacuum of $1\times10^{-9}$ Torr during operation. 

With the compact package we perform Sr MOT trapping to characterize the overall system towards the goal of operating an optical clock. Figure \ref{fig:fig6} (c) and (d) present our characterization of the system's performance to cool and trap strontium atoms in a blue MOT. We notice that a slowing beam is no longer needed to achieve high atom number in the blue MOT, as the high-speed thermal atom flux is replaced by low-speed atom vapor emitted close to the MOT trapping region. The other parameters of blue MOT trapping are similar to our previous numbers in Fig. \ref{fig:fig3}: 6-beam optical intensity $I_{\rm{MOT}}$ = \SI{30}{mW\per cm\squared}, $\Delta_{MOT} \approx \; -1.5~\Gamma$ or $-2\pi$ $\times$ \SI{50}{\mega\hertz}, and magnetic field gradient \SI{50}{G\per\centi\meter}. Typically the system traps \SI{E+6}{} bosonic $^{88}$Sr and \SI{E+5}{} fermionic $^{87}$Sr atoms in half a second of loading time (Fig. \ref{fig:fig6}c). Isotope shift in the transition frequency is accounted by adjusting local oscillator frequency $\left|f_{\rm{LO}}^{\text{(88)}}-f_{\rm{LO}}^{\text{(87)}}\right| \approx$ 30 MHz for the \SI{922}{nm} laser frequency stabilization. 

We notice a strong dependence of the atom vapor pressure and MOT atom number on the operating conditions of atom source laser, which we attribute to the close proximity of atom source to trapping region, and the absence of a differential pumping. The MOT lifetime is also strongly influenced, depicted in the loading curves for $^{88}$Sr (closed circles) and $^{87}$Sr (open circles) shown in Fig. \ref{fig:fig6}(d) under different vacuum pressures from laser diode current at 0.7 A and 0.5 A. In typical operation, \SI{1}{\watt} source laser optical power is generated with 0.5 A laser diode current, and we obtain vacuum pressure of \SI{1E-9}{Torr} with loading curve time constant of \SI{200}{ms} (dashed lines). Increasing the source laser diode current increases the trapped atom number (solid lines), but at the expense of a shorter loading time constant. The loading time constant sets a upper bound on the vacuum lifetime, a critical parameter for many atomic physics applications. We are investigating its effect on clock interrogation of lattice trapped atoms, as we perform narrowline 689 nm cooling of strontium with similar MOT beam generation demonstrated in our previous work \cite{ferdinand_laser-cooling_2025}.

\section{Conclusion}

We have demonstrated a scalable infrastructure for strontium optical clocks that leverages integrated photonics for beam delivery, laser stabilization, and compact system design. Using metasurface optics, we generated complex MOT beam geometries and realized trapping of all stable Sr isotopes with atom numbers commensurate with natural abundances, including efficient trapping of fermionic $^{87}$Sr. Photonic-chip supercontinuum enabled frequency stabilization across visible and near-IR transitions, while a compact second-generation system validated robust operation with reduced size, weight, and power. The final engineered system integrates these advances into a transportable, power-efficient package capable of immediate MOT operation after relocation. These results highlight how integrated photonics can replace bulk free-space optics and establish a versatile platform for optical clocks, with direct applicability to quantum sensing and quantum information.

This research has been funded by the DARPA A-PhI program, AFOSR FA9550-20-1-0004 Project Number 19RT1019, NSF Quantum Leap Challenge Institute Award OMA – 2016244, DARPA LUMOS (HR0011-20-2-0046), and NIST. Trade names provide information, not an endorsement. This work is not subject to copyright in the US.

\bibliography{main_refs_only}
\end{document}